\documentclass[%
 reprint,
 amsmath,amssymb,
prb,
]{revtex4-1}

\usepackage{graphicx}% Include figure files
\usepackage{dcolumn}% Align table columns on decimal point
\usepackage{bm}% bold math
\begin{document}

\preprint{APS/123-QED}

\title{Recurrent approach to effective material properties with \\ application to anisotropic binarized random fields}% Force line breaks with \\

\author{Marc A. Gal\'i}
\author{Matthew D. Arnold}%
\email{Matthew.Arnold-1@uts.edu.au}
\affiliation{%
 School of Mathematical and Physical Sciences, University of Technology, Sydney.
}%


%  \homepage{http://www.Second.institution.edu/~Charlie.Author}
% \affiliation{
%  Second institution and/or address\\
%  This line break forced% with \\
% }%
% \affiliation{
%  Third institution, the second for Charlie Author
% \affiliation{%
%  Authors' institution and/or address\\
%  This line break forced with \textbackslash\textbackslash
% }%

% \collaboration{CLEO Collaboration}%\noaffiliation

\date{\today}% It is always \today, today,

\begin{abstract}
Building on the foundation work of \citet{Brown1955,Milton1981,Milton2002} and \citet{Torquato1985a,Torquato2002}, we present a tractable approach to analyse the effective permittivity of anisotropic two-phase structures. This methodology accounts for successive dipolar interactions, providing a recurrent series expansion of the effective permittivity to arbitrary order. Within this framework, we also demonstrate a progressive method to determine tight bounds that converge towards the exact solution. We illustrate the utility of these methods by using ensemble averaging to determine the micro-structural parameters of anisotropic level-cut Gaussian fields.  We find that the depolarization factor of these structures is equivalent to that of an isolated ellipse with the same stretching ratio, and discuss the contribution of the fourth order term to the exact anisotropy.
\end{abstract}

%Use showkeys class option if keyword
\maketitle

%\tableofcontents

\section{\label{sec:Intro}Introduction}

The effective transport properties of composites have been studied for many years, and a wide range of numerical approaches and results have been surveyed in some excellent books \cite{Milton2002,Torquato2002}. However, some cases remain that have not been thoroughly explored.  Here, we are motivated to develop a better understanding of the effective permittivity tensor of two-phase random anisotropic structures, that are observed for example in obliquely deposited thin films \cite{Walsby2005,Tai2017}. We now briefly outline some of the successes and limitations of the most relevant articles outlined in \citet{Torquato2002}.  

A common approach is to separate constituent phase properties from simple geometric factors that can be used to successively estimate the effective response.   First-order effective-medium approaches are well-known but fail to capture important detail, so much subsequent work has focused on how to incorporate higher order microstructural information.  In particular, Brown \cite{Brown1955} developed a series expansion for isotropic two-phase media using multipoint geometric correlation factors, and gave an explicit integral for a third order microstructural parameter. Teubner and Roberts \cite{Teubner1991} studied a model to deal with three-dimensional isotropic random systems (level-cuts of Gaussian fields). They used integration to find the third order parameter, and averaged direct solutions to find the effective conductivity.  Sen and Torquato \cite{Sen1989} extended Brown's method to anisotropic structures and discussed how to bound the exact results using Pad\'e approximants, with explicit expressions up to 4th order.  

Such bounds have also been a subject of much study \cite{Miller1969,Milton1981,Bergman1981,Milton1982,Torquato1985a,Torquato1991,Torquato1983,Torquato1990,Torquato2002,Milton2002}, because they progressively narrow the possible results by incorporating increasing level of microstructural parameters (MSP). For anisotropic structures, it is informative to plot bounds in the \emph{“G-closure”} permittivity-component space popularized by Cherkaev \cite{Cherkaev1996}. In this space, bounds can be simple rectangles (e.g. the well-known Wiener bound \cite{Wiener1912}), or optimally are curved.  Optimal bounds are known for low-order: for example the bounds independently discovered by Tartar \cite{Murat1994} and Cherkaev \cite{Cherkaev1996} only require fill factor information.  We note that more recently, Engstr\"{o}m \cite{Engstrom2005} demonstrated how internal parameters in Torquato’s framework are themselves bounded by previous orders, giving explicit expressions up to 4th order.  
% * <matthew.arnold-1@uts.edu.au> 2018-08-31T04:19:17.043Z:
%
% > study
%
% ^.

However, high order bounds on permittivity were not illustrated, which we will address in this article.  More generally, the process of deriving high order results becomes increasingly complicated, and determining numerical results becomes intractable due to the need to calculate high-dimension correlation functions and then integrate them over these dimensions.  In this article, we reframe the aforementioned works to show how to efficiently determine anisotropic geometric factors to high order and demonstrate how to generate tight bounds that constrain the anisotropy. To exemplify this process and address our original aim, we extend beyond the results of Roberts and Teubner to investigate the effective permittivity tensor of two-dimensional anisotropic ensembles.

%---------------------------------------------
%\subsection{\label{ssec:Theoretical Framework}Theoretical Framework}

%---------------------------------------------
\section{Theoretical Framework}
The effective permittivity of a structure may be determined at sufficiently small sizes by averaging the induced polarization field.  The region of validity has been thoroughly reviewed elsewhere \cite{Arnold2017}, including the applicability the effective medium approach to wire metamaterials \cite{Alu2011,Simovski2012}.  The determination of the induced field has been extensively studied, in particular by finding the solution of a self-consistent electrodynamic dipole system \cite{Draine1994,Draine2008}. For simplicity we will consider the electrostatic limit of this which can be written as: 
\begin{equation}
\vec{F}(\vec{r}) = \vec{E_0}(\vec{r}) + \int \tilde{\mathbb{G}}(\vec{r},\vec{s}) \left[ \epsilon (\vec{r})-\epsilon_i \right] \vec{E}(\vec{s}) d\vec{s},
\label{eq:Fsol}
\end{equation}
where $\vec{F}$ is the \emph{cavity intensity field}, also known as \emph{Lorentz field}, $\vec{E_0}$ is the incident electric field and $\epsilon(\vec{r})$ is the dielectric function. We defined the interaction tensor as
\begin{equation}
\tilde{\mathbb{G}}(\vec{r},\vec{s}) = \frac{d\cdot \hat{\rho}\hat{\rho} -  \mathbb{I}}{2\pi(d-1) \epsilon_i \rho^{d} }.
\label{eq:interactionTensor}
\end{equation}
where $d$ is the dimensionality and $\rho \equiv |\vec{r}-\vec{s}|$ is the distance between $\vec{r}$ and $\vec{s}$, with its unit vector being $\hat{\rho}\equiv \frac{\vec{r}-\vec{s}}{|\vec{r}-\vec{s}|}$, and $\mathbb{I}$ is the identity matrix.

Finding the self-consistent solution is slow and so there has been extensive research effort on how to estimate the solution by incorporating increasing structural information. Notably, by successive substitution of the self-consistent relationship a Taylor series representation of the isotropic \cite{Brown1955} and anisotropic \cite{Sen1989,Torquato1990} permittivity is found:

\begin{equation}
\frac{\hat{\epsilon}_{\text{eff}}}{\epsilon_i}=\sum_{n=0}^{\infty} \hat{a}_n \chi^n. 
\label{eq:effepsExp}
\end{equation}

This (and similar series) glibly separates the contrast ratio ($\chi$) of the phases  from geometric parameters ($\hat{a}_n$), allowing useful insights into the effect of geometry. To determine the $\hat{a}_n$ parameters we can start from the following transformed series \cite{Sen1989}:

\begin{align}
(\beta_{ji}\phi_j)^2 <\beta_{i}>^{-1}  = \phi_j \beta_{ji} \mathbb{I}-\sum_{n=2}^{\infty} \hat{A}_n^{(j)}\beta_{ji}^n,
\label{eq:EffBetaExp}
\end{align}
where $\beta_{ji}=\frac{\epsilon_j-\epsilon_i}{\epsilon_j+(d-1)\epsilon_i}$ is the polarizability due to phases $\epsilon_i$ and $\epsilon_j$, $<\beta_{i}> = \left( \hat{\epsilon}_{\text{eff}}+(d-1)\epsilon_i \mathbb{I} \right) ^{-1} \left( \hat{\epsilon}_{\text{eff}}-\epsilon_i \mathbb{I} \right)$ is the effective polarizability tensor of the composite, $\phi_j$ is a fill-factor, and the tensor coefficients $\hat{A}_n^{(j)}$ were introduced by \citet{Torquato1985a} (see also \cite{Sen1989,Torquato1990}) which are equivalent to the tensors introduced by \citet{Brown1955}.

\begin{align}
\hat{A}_2^{(j)} &= \kappa \int \left[ p_2^{(j)}(\vec{r}_1,\vec{r}_2) - \phi_j^2 \cdot \mathbb{I} \right] \tilde{\mathbb{G}}_{12} d\vec{r_2} \label{eq:A2_def} \\ 
\hat{A}_n^{(j)} &= (-1)^n\phi_j^{2-n}(\kappa)^{n-1} \int \dots \int d\vec{r}_2 d\vec{r}_3 \dots d\vec{r}_n \\
& \times \tilde{\mathbb{G}}_{12}\cdot \tilde{\mathbb{G}}_{23}\cdots \tilde{\mathbb{G}}_{n-1 n}\cdot C_n^{(j)}, \label{eq:An_def}
\end{align}
where $\kappa$ is $\epsilon_i k_0^2 d$ with $k_0$ the wavenumber, and $C_n^{(j)}$ is the determinant of the \emph{n-probability function matrix} of phase $j$:

\begin{equation}
C_n^{(j)} =  
\begin{array}{|c c c c c|}
p_2^{(j)} & p^{(j)}_1 & 0 & \dots & 0 \\
p^{(j)}_3 & p^{(j)}_2 & p^{(j)}_1 & \dots & 0 \\
\vdots & \vdots & \vdots & \dots & \vdots \\
p^{(j)}_{n-1}& p^{(j)}_{n-2} & p^{(j)}_{n-3} & \dots  & p^{(j)}_1 \\
p^{(j)}_n& p^{(j)}_{n-1} & p^{(j)}_{n-2} & \dots  & p^{(j)}_2 
\end{array}\quad .\nonumber
\end{equation}

Here, the structural parameters $\hat{A}_n^{(j)}$ (which are needed to find $\hat{a}_n$) have been written in terms of multipoint correlation factors $p^{(j)}_n$, but the calculation beyond 2nd order is increasingly difficult, since they must be determined by multidimensional integration.  A similar problem occurs in perturbation expansion of the electromagnetic scattering of rough surfaces \cite{ODonnell2001}, so alternative approaches to evaluating such expressions are of significant value. In this work we present a more tractable approach to obtain high order parameters for the electrostatic effective medium problem, by a careful arrangement of previous derivations, accounting for successive dipole interactions.

%----------------------------------------------
\section{Successive Interaction Approach} \label{sec:SuccAppr}

Following previous work by Brown \cite{Brown1955}, we go back to the successive substitution of the interaction equation

\begin{widetext}
\begin{align}
\vec{P}(\vec{r_1}) =&  \kappa \beta_{ji}\mathcal{I}^{(j)}(\vec{r_1}) \cdot \left( \vec{E_0} + \int d\vec{r_2} \tilde{\mathbb{G}}(\vec{r_1},\vec{r_2}) \left( \kappa \beta_{ji}\mathcal{I}^{(j)}(\vec{r_2}) \cdot \left(\vec{E_0} + \int d\vec{r_3}  \tilde{\mathbb{G}}(\vec{r_2},\vec{r_3}) ( \dots ) \right) \right) \right), \nonumber \\
=& \kappa \beta_{ji} \mathcal{I}_1^{(j)}\vec{E_0} +   (\kappa \beta_{ji})^2 \int  \mathcal{I}_1^{(j)}\mathcal{I}^{(j)}_{2}\tilde{\mathbb{G}}_{12}\cdot \vec{E_0}  d\vec{r_2} \nonumber \\
& +  (\kappa \beta_{ji})^3 \int d\vec{r_2} \int \mathcal{I}_1^{(j)}\mathcal{I}_2^{(j)}\mathcal{I}_3^{(j)}\tilde{\mathbb{G}}_{12}\cdot \tilde{\mathbb{G}}_{23}\cdot \vec{E_0} d\vec{r_3} + \dots,
\label{eq:P1stIteration}
\end{align}
\end{widetext}
where to simplify notation we have defined the interaction tensor \\ $\tilde{\mathbb{G}}_{ij}\equiv \tilde{\mathbb{G}}(\vec{r_i},\vec{r_j})$ and the indicator (or characteristic) function $\mathcal{I}_n^{(j)}\equiv \mathcal{I}^{(j)}(\vec{r_n})$.

However, rather than rewriting the geometry products in terms of the multipoint correlation functions, we leave the successive interactions intact during averaging, giving an explicit power series in the geometry-interaction product:
\begin{align}
<\vec{P}> =&  \vec{E}_0 \kappa \beta_{ji} <\mathcal{I}^{(j)}>  + \vec{E}_0 \left( \kappa \beta_{ji} \right)^2< \mathcal{I}^{(j)} \cdot \tilde{\mathbb{G}} \cdot \mathcal{I}^{(j)} > \nonumber \\
& + \vec{E}_0 \left( \kappa \beta_{ji} \right)^3< \mathcal{I}^{(j)} \cdot \tilde{\mathbb{G}} \cdot \mathcal{I}^{(j)} \cdot \tilde{\mathbb{G}} \cdot \mathcal{I}^{(j)} > + \dots ,
\label{eq:P1stIterationMC}
\end{align}
This expression can be written as a series expansion on $\left( \kappa \beta_{ji} \right)$,
\begin{align}
<\vec{P}> =& \vec{E}_0 \sum_{n=1}^{\infty} \underbrace{< \mathcal{I}^{(j)} \prod_{i=2}^n \left( \tilde{\mathbb{G}} \cdot \mathcal{I}^{(j)} \right) > }_{q_n} \left( \kappa \beta_{ji} \right)^n, 
\label{eq:SerExpPMC} \\
\nonumber \\
q_n =& < \mathcal{I}^{(j)} \prod_{i=2}^n \left( \tilde{\mathbb{G}} \cdot \mathcal{I}^{(j)} \right) >. \label{eq:qn}
\end{align}
Similar approaches are alluded by Milton \cite{Milton2002} and Torquato \cite{Torquato2002}, but they do not seem to have been fully developed in the framework shown here.
Following a similar process to Brown, we invert the series to expand in terms of the averaged polarization field, and then remove the incident field dependence, giving an equivalent series expansion to Torquato:

\begin{align}
(\beta_{ji}\phi_j)^2 <\beta_{i}>^{-1}  = \phi_j \beta_{ji}\mathbb{I} -\sum_{n=2}^{\infty} \mathbb{A}_n^{(j)}\beta_{ji}^n,
\label{eq:EffBetaExpMC}
\end{align}
where the tensors $\mathbb{A}_n$ are the equivalent $\hat{A}_n$ tensors obtained using successive interactions (instead of using $p_n$).

\begin{subequations}\label{eq:IterAn}
\begin{align}
\mathbb{A}_2^{(j)} &= \kappa \left( < \mathcal{I}^{(j)} \cdot  \tilde{\mathbb{G}} \cdot \mathcal{I}^{(j)} > - \phi_j^{2}<\tilde{\mathbb{G}}> \right)  \\ 
\mathbb{A}_n^{(j)} &= (-1)^n\phi_j^{2-n}(\kappa)^{n-1} r_n^{(j)}, 
\end{align}
\end{subequations}
where $r_n^{(j)}$ can be expressed as a determinant of the $q_n$ coefficients:

\begin{equation}
r_n^{(j)} =  
\begin{array}{|c c c c c|}
q_2 & q^{-1}_1 & 0 & \dots & 0 \\
q_3 & q_2 & q^{-1}_1 & \dots & 0 \\
\vdots & \vdots & \vdots & \dots & \vdots \\
q_{n-1}& q_{n-2} & q_{n-3} & \dots  & q^{-1}_1 \\
q_n & q_{n-1} & q_{n-2} & \dots  & q^{-1}_2 
\end{array}\quad ,\nonumber
\end{equation}
where to simplify notation we have omitted the phase dependence on $q_n^{(j)}$ (implicit on the definition of $q_n$ in equation \ref{eq:SerExpPMC}).
This greatly improves the tractability of $\mathbb{A}$ since each interaction-geometry power can be determined recursively from the previous, which can be efficiently implemented using a convolution theorem.

Using the definition of the $\mathbb{A}_n$ tensors we can obtain the coefficient tensors $\hat{a}_n$ of the series expansion of the effective permittivity (eq. \ref{eq:effepsExp}), but instead of solving explicitly for individual orders, we write this in general form allowing determination to any order

\begin{subequations}
\label{eq:XiDef}
\begin{align}
\hat{\xi}_0 &= 0,\\
\hat{\xi}_1 &= \phi_j \mathbb{I},\\
\hat{\xi}_2 &= d^{-1} \left( \phi_j \mathbb{I}- \hat{A}^{(j)}_2 \right),\\
\hat{\xi}_n &= d^{1-n} \sum_{k=3}^{n} \binom{n-3}{k-3} (-1)^{n+k+1} \hat{A}_k^{(j)} , \qquad  \left( \forall n>2 \right) .
\end{align}
\end{subequations}
\begin{subequations}\label{eq:chi_to_qnpn}
\begin{align}
&\hat{q}_n=\hat{p}_n=\hat{\xi}_n, \qquad \left( \forall n\neq 2 \right),\\
&\hat{q}_2= \hat{\xi}_2 -\frac{\phi_j^2}{d}\mathbb{I}, \\
&\hat{p}_2 = \hat{\xi}_2 +\frac{\phi_j^2}{d}(d-1) \mathbb{I}.
\end{align}
\end{subequations}
\begin{equation}
\hat{a}_n =  \sum_{m=n}^N\hat{p}_m \cdot \hat{q}_{m-n}^{-1},
\label{eq:an_def}
\end{equation}
where $N$ is the truncation order. From this point, in order to simplify notation we will define $\chi = \frac{\epsilon_1}{\epsilon_2}-1$ with $\epsilon_2>\epsilon_1$. Then with these definitions of $\hat{a}_n$ and $\xi$ we can express the relative effective permittivity as a power series expansion (eq. \ref{eq:effepsExp}).

%-----------------------
\section{High Order Bounds: Tight Bounds}

With the methodology presented in the previous section we can obtain higher orders of the series expansion of the effective permittivity. Consequently we can use the methodologies presented by other authors \cite{Milton2002,Torquato2002} to obtain the bounds of the effective permittivity. In particular Torquato transformed the Taylor series to a Pad\'e series which are bounds on the Taylor series. \citet{Engstrom2005} clarified this via an inverse series for one of the bounds, and also demonstrated that the $\hat{a}_n$ tensor parameters are themselves bounded by the previous order.

\begin{figure}
 \centering \includegraphics[scale=.35]{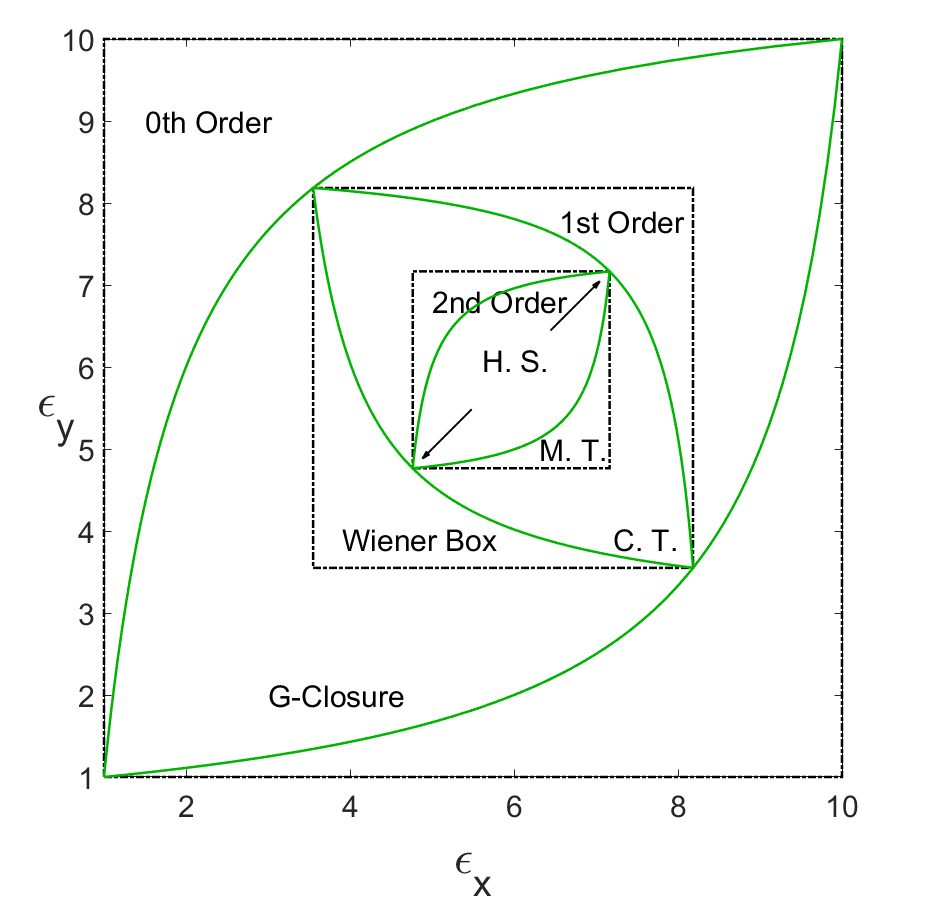}
 \caption{In this figure several known bounds in the literature can be seen (Wiener Box \cite{Wiener1912}, \emph{C.T.} stands for Cherkaev \cite{Cherkaev1996} and Tartar \cite{Murat1994}, \emph{H.S.} for \citet{Hashin1962}, \emph{M.T.} for \citet{Milton2002,Torquato2002}). These bounds have been obtained for a two-phase 2D material described by a square unit cell with a circle, where the occupancy phase has $\epsilon_1 = 10 $ and fill factor of $80\%$, and the background phase has $\epsilon_2 = 1$ and fill factor of $20\%$.}
 \label{fig:BoundsDescription}
 \end{figure}

At this point, we illustrate the nature of bounds by considering their shape in permittivity-component space (figure  \ref{fig:BoundsDescription}), a representation popularized by \citet{Cherkaev1996}. The most obvious bound is the fill-factor dependent Wiener rectangle \cite{Wiener1912} with corners corresponding to parallel plate geometries, but the rectangle is not optimal. Instead, the track of the corners over the allowed range of fill factors generates the so-called \emph{G-closure} which is the outermost optimal bound, independent of geometry. Cherkaev and independent authors \cite{Cherkaev1996} determined the next optimal bound, which intersects with the isotropic Hashin-Strikman points \cite{Hashin1962}, and only depends on fill-factor. A naive reading of Torquato's framework would generate rectangular (non-optimal) bounds, but by a relatively simple extension of Engstr\"{o}m’s work we can determine a general algorithm to obtain arbitrarily high orders of tight bounds on permittivity.  Due to the coincidence with low order optimal bounds, we conjecture that the entire series is optimal.

We determine the extremal values of the curved bounds, starting from Engstr\"{o}m's method of bounding $\hat{a}_n$ by imposing equality between the rectangular bounds of consecutive order at their extremal points, noting that the tensor coefficient $ \hat{a}_n $ or $ \hat{\alpha}_n $ of the higher order bound are considered tensor parameters $\{ \hat{a}_n \}$ or  $\{ \hat{\alpha}_n \}$ (are the variables of the equation). Thus the solution of this equalities are the extremal points of the tensor parameters, i.e. $\lfloor \hat{a} \rfloor$ and $\lceil \hat{a} \rceil$ .

\begin{equation}
\begin{cases}
\lceil \hat{\epsilon} \rceil_{n} &= \lceil \hat{\epsilon} \rceil_{n+1}(\left\{ \hat{a}_{n+1}\right\}), \\
\lfloor \hat{\epsilon} \rfloor_{n} &= \lceil \hat{\epsilon} \rceil_{n+1}(\left\{ \hat{\alpha}_{n+1}\right\}).
\end{cases}
\label{eq:ext_bound_an}
\end{equation}

Thus, the values that satisfy equations \ref{eq:ext_bound_an} ($\lfloor \hat{a}_{n+1} \rfloor$  and $\lceil \hat{a}_{n+1} \rceil$) are the extreme values of the leaf bounds (the vertices of the leaf):
\begin{align}
\lfloor \hat{a}_{n+1} \rfloor \leq \left\{ \hat{a}_{n+1}\right\} \leq \lceil \hat{a}_{n+1} \rceil,
\label{eq:extremal_an}
\end{align}
equivalently the tensor parameters $\{ \hat{\alpha}_n \}$, and extreme points $\lfloor \hat{\alpha} \rfloor$ and $\lceil \hat{\alpha} \rceil$ can be also determined.

Further, we can now express the optimal permittivity bounds of any order in the following way:
\begin{subequations}\label{eq:LeafBAlgo}
\begin{align}
\lceil \hat{\epsilon} \rceil_n^{LB} &= \epsilon_2 \left( \sum_{k=0}^n \hat{a}_k \chi^k + \left\{ \hat{a}_{n+1} \right\} \chi^{n+1} \right) , \\
\left( \lfloor \hat{\epsilon} \rfloor_n^{LB} \right)^{-1} &= \frac{1}{\epsilon_1} \left( \sum_{k=0}^n \hat{\alpha}_k \chi^k + \left\{ \hat{\alpha}_{n+1} \right\} \chi^{n+1} \right),
\end{align}
\end{subequations}
where to trace the bounds we need to sweep  $\left\{ \hat{a}_{n+1} \right\}$ and $\left\{ \hat{\alpha}_{n+1} \right\}$ through the range defined in equations \ref{eq:extremal_an} and \ref{eq:LeafBAlgo} respectively.  Mapping of curved bounds is implicit in some previous works, for example in the context of complex isotropic permittivity \cite{Milton2002,Engstrom2005}, but to the best of our knowledge has not been illustrated for high order anisotropy.

\begin{figure}
 \centering \includegraphics[scale=.22]{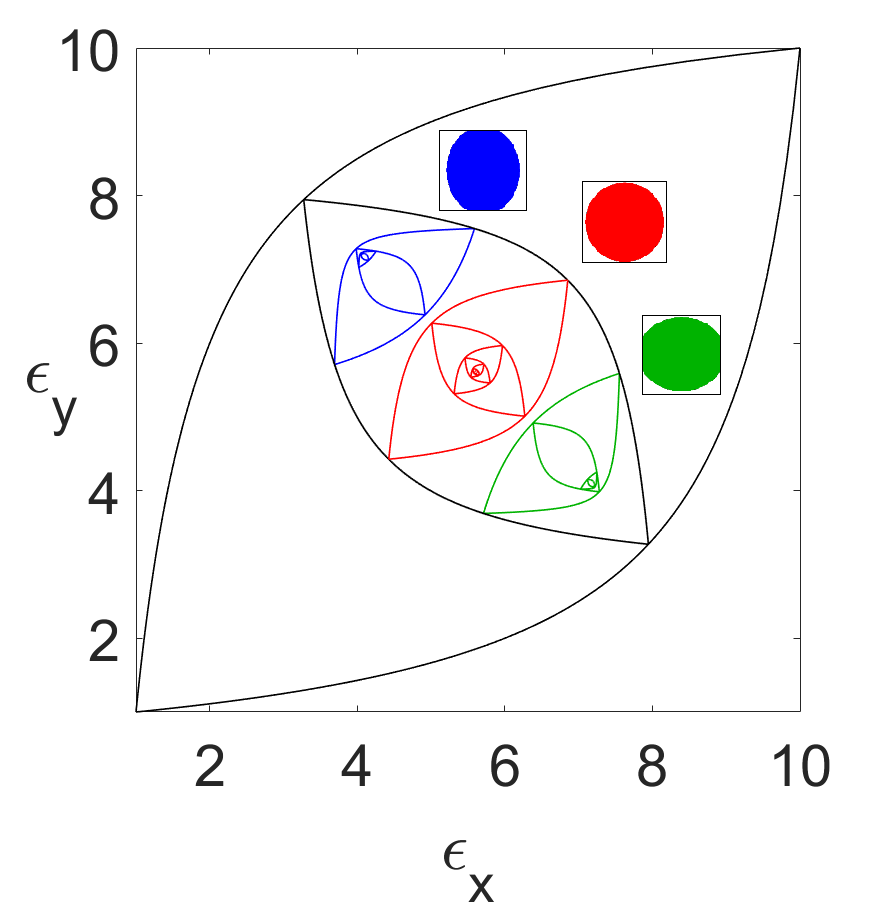}
 \includegraphics[scale=.17]{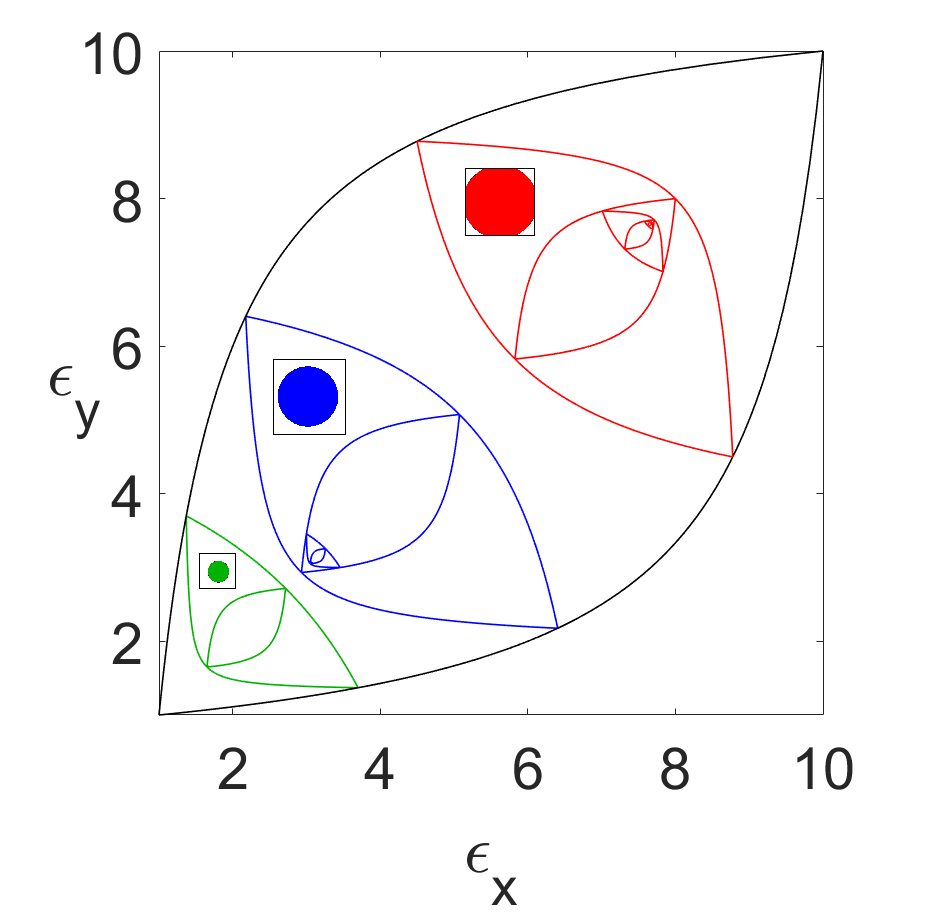}
 \caption{In these figures we show the high order leaf bounds for different structures. Left image depicts high order leaf bounds for structures with the same fill factor (80$\%$) and different anisotropy ratios. Right image shows high order leaf bounds for isotropic unit cells, but different fill factors (30$\%$, 60$\%$ and 90$\%$). In these calculations the occupancy phase has $\epsilon_1 = 10 $ and the background phase has $\epsilon_2 = 1$.}
 \label{fig:BoundsAnisFF}
 \end{figure}

The convergence with order of the curved bounds gives information on the level of anisotropy or connectivity (fill factor) of the structure. Thus in figure \ref{fig:BoundsAnisFF} we can see the different distribution of the bounds changing the anisotropy (left) or changing the fill factor (right). For both figures black bounds represent the common bounds of the different structures. Thus structures with same fill factor share the zeroth and first bounds, while structures with different fill factor (right image in figure \ref{fig:BoundsAnisFF}) only share the zeroth order bound. From this figure we can see that high order bounds quickly converge to the solution in structures with high or low fill factor, or high anisotropy. Therefore high order bounds are most interesting for structures that are near percolation (fill factors around $50\%$) and nearly isotropic.

%------------------------------------------
\section{Micro-Structural Parameters}

Micro-Structural Parameters (MSP) are commonly used in the literature to encapsulate meaningful geometry-dependent information. In this work we define the MSP of arbitrary order using the tensor coefficient  $\hat{a}_n$ presented in the previous section. The MSP of n-order are a normalization of the coefficient tensors using the extreme points of the leaf bounds.

\begin{equation}\label{eq:MSPdef}
\hat{\Upsilon}_n^{(j)}  = \frac{\hat{a}_n- \lfloor\hat{a}_n\rfloor }{\lceil \hat{a}_n \rceil- \lfloor \hat{a}_n \rfloor} ,
\end{equation}
where $\hat{\Upsilon}_n^{(j)}$ is the micro-structural tensor of n-order for phase $j$, $\hat{a}_n$ is the tensor of the series power expansion of order n (eq. \ref{eq:effepsExp}), and $\lceil \hat{a}_n \rceil$, $\lfloor \hat{a}_n \rfloor$ are the maximum and minimum values that $\hat{a}_n$ can achieve on the leaf bounds of n-order (see equations \ref{eq:ext_bound_an} and \ref{eq:extremal_an}). 

This definition mostly agrees with the definitions used by other authors, who investigated MSP of lower orders. Specifically:
%\\
%$\hat{\Upsilon}^{(j)}_1 = \hat{\phi}_j$ is the fill factor.\\
%$\hat{\Upsilon}^{(j)}_2 = \hat{L}$ is the depolarization tensor.\\
%$\hat{\Upsilon}^{(j)}_3 = \hat{\zeta}$ gives information on the connectivity. \\
%$\hat{\Upsilon}^{(j)}_4 \sim \hat{\gamma}$ gives high order information on anisotropy. 
\begin{align*}
    &\hat{\Upsilon}^{(j)}_1 = \hat{\phi}_j, \\
    &\hat{\Upsilon}^{(j)}_2 = \hat{L}, \\
    &\hat{\Upsilon}^{(j)}_3 = \hat{\zeta}, \\
    &\hat{\Upsilon}^{(j)}_4 \sim \hat{\gamma}.
\end{align*}
The first two parameters are obviously the fill-factor and depolarization tensor respectively.  As evident in the literature and in figures  \ref{fig:IsoValidation} and \ref{fig:EllipseValidation}, the high order parameters are prominent for non-smooth shapes (e.g. with sharp edges or percolation), even orders being associated with anisotropy.  We note a trivial difference between definitions at fourth order: with isotropy $\hat{\gamma}\rightarrow 0$, but our micro-structural parameter  $\hat{\Upsilon}_4 \rightarrow 1/2$.  

As the MSP are tensors, for anisotropic systems we have different values in each main direction. In order to compare isotropic and anisotropic systems and their MSP of different orders, in two dimensions we find convenient to define the following new quantities to analyze even orders ($\Delta \hat{\Upsilon}$) and odd orders ($<\hat{\Upsilon}>$):
\begin{align}
<\hat{\Upsilon}_{n}> &= \frac{\left( \hat{\Upsilon}_n \right)_{11}+ \left( \hat{\Upsilon}_n \right)_{22}}{2}, \quad \text{for n odd} \label{eq:AvUps}\\
%& \\
 \Delta \hat{\Upsilon}_n &= \left( \hat{\Upsilon}_n\right)_{22}- \left(\hat{\Upsilon}_n\right)_{11},\quad \text{for n even}\label{eq:DeltUps}
\end{align}
with this notation isotropic structures yield $\Delta \hat{\Upsilon}_{2n} = 0$.
%------------------------------------------
\section{Validation}
\begin{figure}
 \centering \includegraphics[scale=.58]{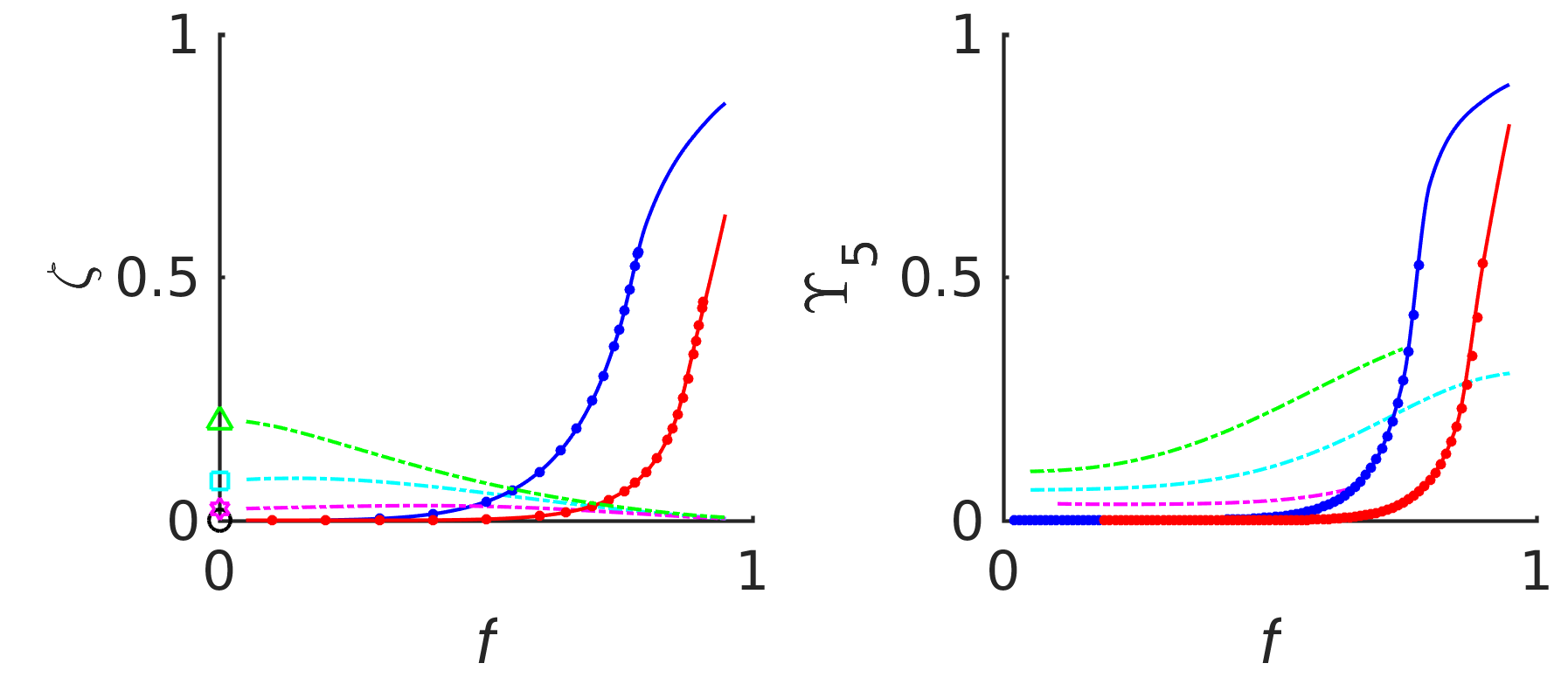}
 \caption{(Color online) Validation of odd microstructural parameters of regular isotropic configurations.  Dashed lines (cyan, magenta and light green) are squares, hexagons and triangles respectively in their natural lattices, with markers from \citet{Hetherington1992}.  Solid (blue and red) lines correspond to discs in square and close-packed lattices respectively distinguished by their percolation thresholds, with dots representing results from \citet{McPhedran1981} (fifth order were inferred from $Q_5$).  Lines shown are consistent to better than 0.01 on phase interchange. } 
 \label{fig:IsoValidation}
 \end{figure}

\begin{figure}
 \centering \includegraphics[scale=.58]{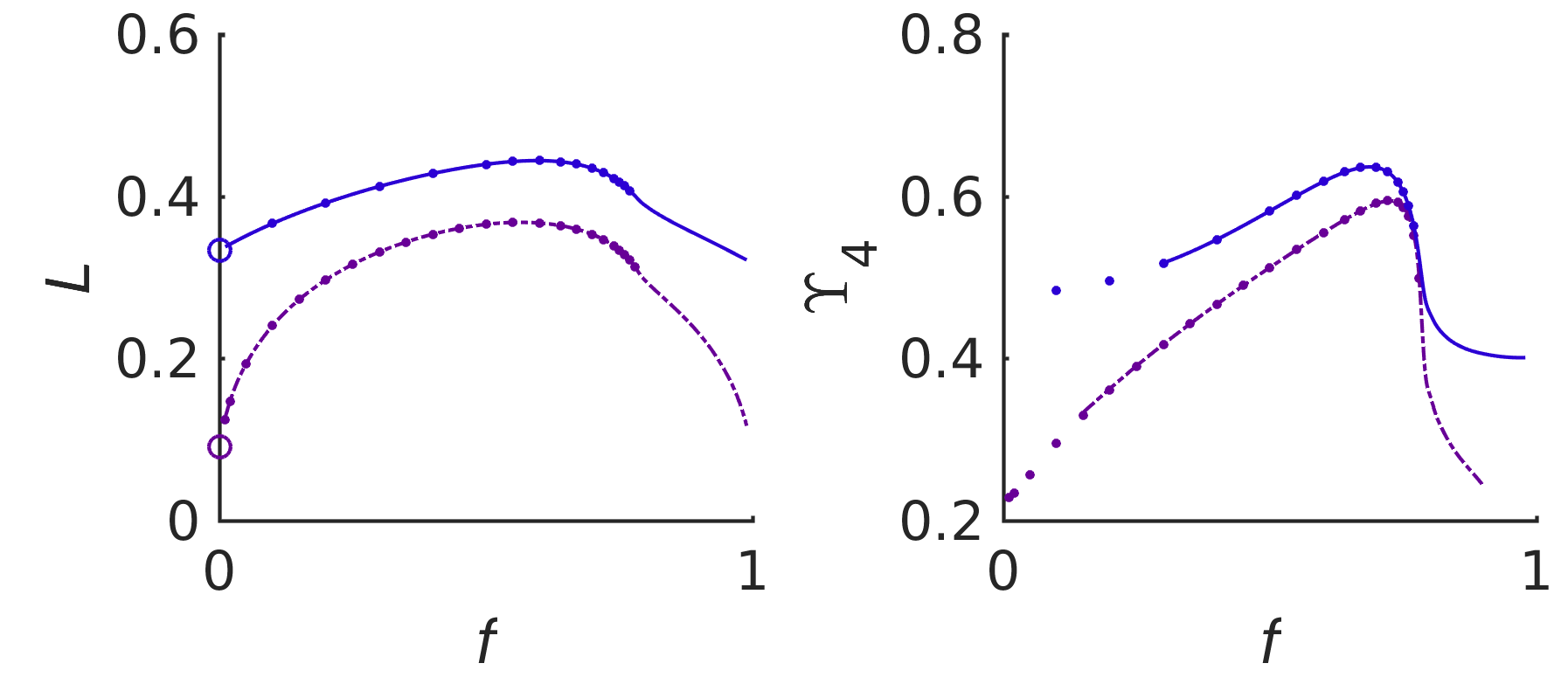}
 \caption{(Color online) Validation of even microstructural parameters of ellipses in rectangular arrays. The lines are for stretched isotropic arrays (i.e. equal lattice and semi-axis ratios): ratios 10 (dashed) and 2 (solid). Results shown are consistent to better than 0.01 on phase interchange. The dots were derived from an independent Rayleigh expansion method  \cite{Yardley1999}.  Circular markers are the expected dilute depolarization.   } 
 \label{fig:EllipseValidation}
 \end{figure}

In order to demonstrate the validity of the method we replicate some tabulated results from the literature, including arrays of circles and polygons (figure \ref{fig:IsoValidation}) and ellipses (figure \ref{fig:EllipseValidation}). We also tested against the well-known phase-inversion theory of \citet{Keller1964}. 

In general we find good agreement for permittivity, provided that sampling is sufficiently fine - near geometric singularities (e.g. near percolation), convergence is somewhat slower. It should be noted that near well-bounded points (e.g. low or high fill-factor), high order microstructural parameters are more difficult to determine accurately, but they have minimal effect on the estimated permittivity at low contrast.  For simple geometries, several methods could provide higher quality estimates of the interaction terms $\hat{A}_n$. Some examples of those are: space-conforming methods (such as the boundary element method \cite{Lu1999}), series expansion methods \cite{McPhedran1980} (such the Rayleigh method  \cite{Nicorovici1996,Yardley1999}) or integral methods \cite{Hetherington1992}. Nonetheless, to study arbitrary random geometries, the discrete dipole sampling is more convenient.

%------------------------------------------
\section{Application to Binarized Random Fields}
Binarized (or level-cut) random fields (BRF) are a useful model for the geometries that we wish to understand.  They are readily created by selection of an appropriate filter kernel or power spectrum which is applied to Gaussian noise and then binarized by choosing an appropriate threshold to achieve the desired fill-factor. The properties of three-dimensional isotropic fields were carefully studied by Roberts and Teubner \cite{Roberts1995,Roberts1999}. Here we consider spatially anisotropic two-dimensional fields, which due to reduction in symmetry would present challenges for previous approaches even at third order.  We consider some representative kernels, investigating the effect of fill factor for isotropic kernels and the anisotropy due to stretching with $\phi$=1/2.

%------------------------------------------}
\subsection{Effect of sampling and expected results}

To obtain valid results using this methodology a few numerical issues must be considered: (i) sampling should be sufficiently fine within the characteristic length scale and (ii) to avoid boundary effects in random structures the unit cell should be sufficiently large compared to the decay length.  To check the validity of our sampling we ran convergence tests increasing both (i) and (ii) in tandem. We used two measures of error: (a) the variation between phase inversions and (b) the convergence to the best estimate (either a known result or best sampling).

It is worth noting that exact results of the MSP are known for some parameters: third order is 1/2 for $\phi$=1/2 \cite{Roberts1995} and even orders are 1/2 for isotropic at any fill factor. We propose one further exact result, which is that the depolarization of kernels of elliptical symmetry equals the depolarization of an equivalent ellipse, with reasoning as follows. The depolarization of dilute ellipses is known to match isolated ellipses. Under elliptical symmetry, the integral of the necessary product around corresponding concentric elliptical rings is simply the integral of the interaction tensor, and this integral can be shown to be zero (e.g. by symmetry). Thus in this frame the depolarization arises entirely from the conditionally-convergent integral of the singularity (within an infinitesimal ellipse).  The correlation function of a physical kernel becomes constant in the limit near the origin, and hence any elliptical kernel must produce exactly the depolarization of an equivalent effective ellipse.

This is consistent with but less obvious than the results found for symmetric-cell (SC) materials.  Since SC have no correlation between cells,  the shape of the autocorrelation function and hence the effective parameters of such structures are the same as an isolated inclusion \citet{Miller1969,Miller1969b}, and this extends to the depolarization of anisotropic SC \cite{Hori1973,Hori1973a}. The BRF structures presented here are more general in that they are correlated over an extended range, and our result applies to an \textit{equivalent} ellipse with aspect ratio that matches kernel contours at all length scales.

Returning to sampling, we found that both the variation on phase inversion and the variation from the expected values decrease as the sampling is increased, with microstructural parameters up to fourth order having error around 0.01 at the sampling we used for our main results (length/pixel and period/length of 30). In general, convergence is degraded at high order, near highly constrained points, and when the kernel makes sampling difficult (e.g. the slow decay of sinc). 

%-----------------------------------------------------------
\subsection{Results and Analysis}
\begin{figure}
\centering \includegraphics[scale=.17]{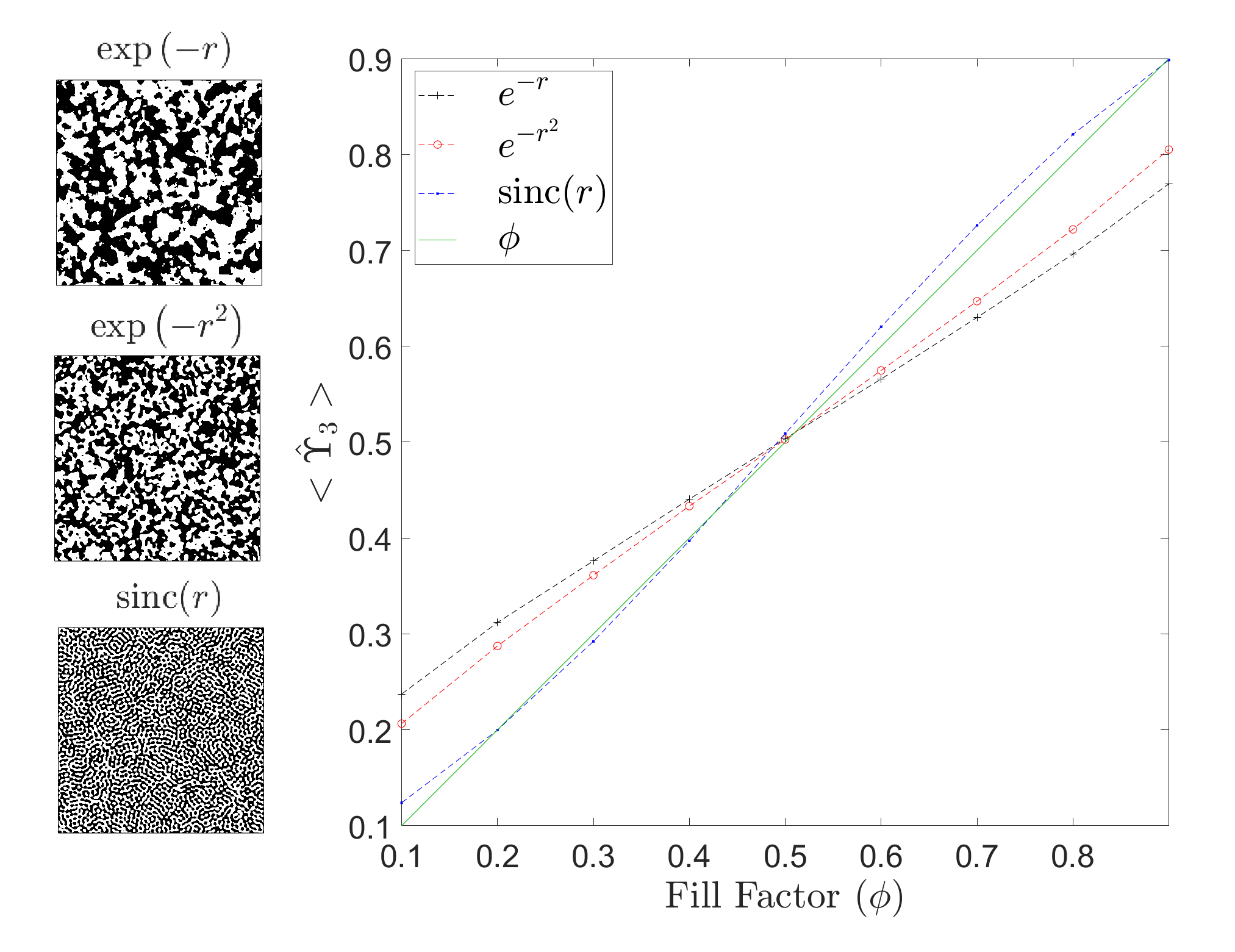}
\caption{(Color online) Study of isotropic systems generated with various kernels. On the left side are shown the unit cell structures for the three different kernels presented here with a fill factor of 50 $\%$. In the plot we show the dependence with fill-factor of the third order micro-structural parameter. These results are obtained averaging 100 realizations and the ratio between the decaying length of the kernel and the unit cell is 1/30.}
\label{fig:IsoMSP34}
\end{figure}

\begin{figure}
\centering \includegraphics[scale=0.2]{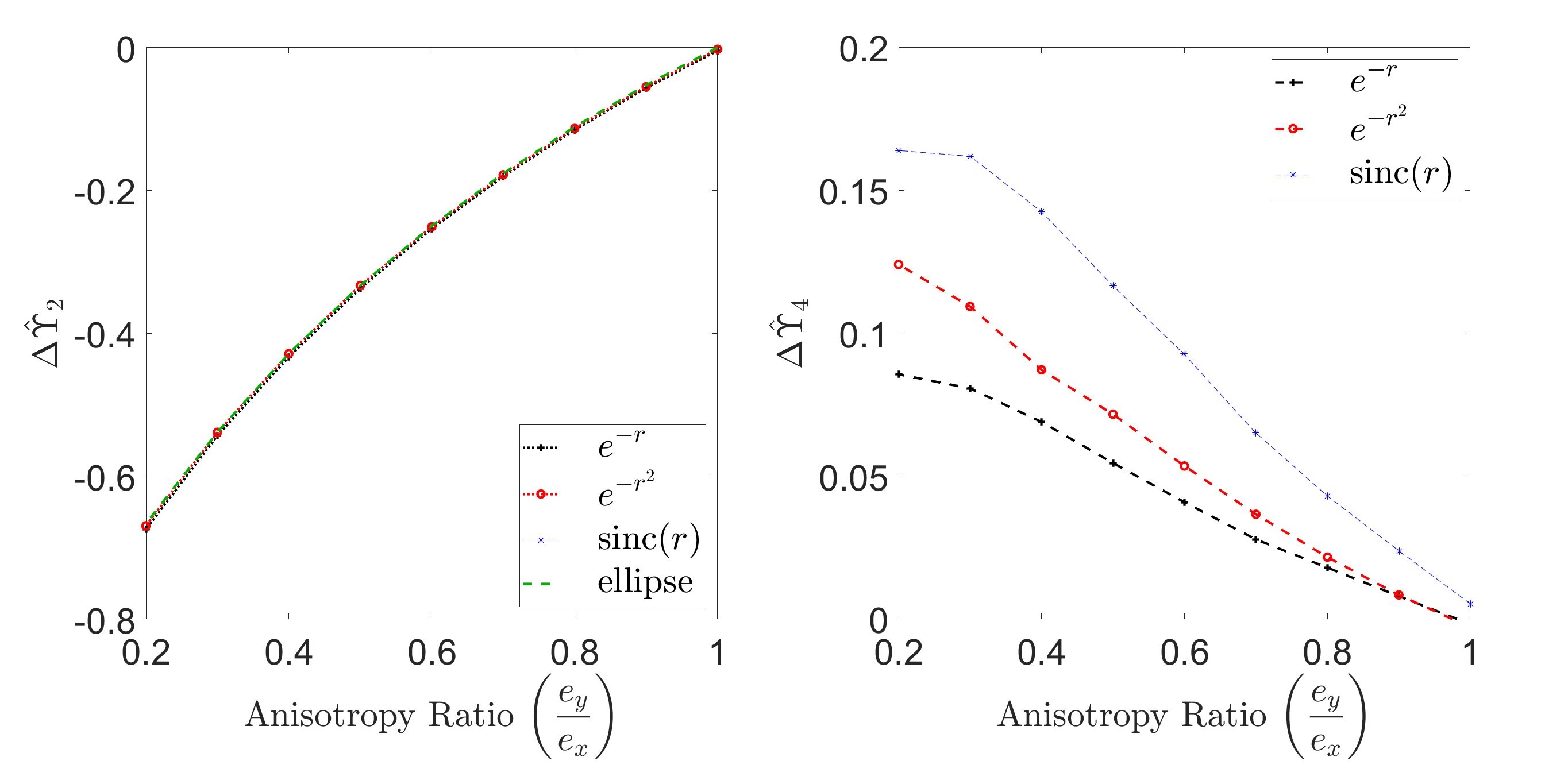}
\caption{(Color online) Study of anisotropic systems generated with exponential, Gaussian and damped sinc kernels. On the left picture we show the anisotropy dependence of the difference of the diagonal elements of the micro-structural parameter of second order ($\Delta \hat{\Upsilon}_2$) and the depolarization factors of an ellipsoid with same degree of anisotropy (green dashed line). The right plot shows the difference between the diagonal elements of the micro-structural parameter of fourth order ($\Delta \hat{\Upsilon}_4$) in terms of the anisotropy ratio. In the plots below we show the unit cell structures for the different kernels at 50 $\%$ fill factor and $e_y/e_x=0.3$ anisotropy ratio. As in figure \ref{fig:IsoMSP34} these results are obtained averaging 100 realizations and the ratio between the decay length of the kernel and the unit cell is 1/30.}
\label{fig:AniMSP34}
\end{figure}

Using isotropic (figure \ref{fig:IsoMSP34}) and anisotropic (figure \ref{fig:AniMSP34}) kernels to generate our structures we can study the relevance of the MSP of odd and even orders depending on the system morphology. In this paper we will show results of the MSP up to fourth order analysing the results using the variables defined in equations \ref{eq:AvUps}-\ref{eq:DeltUps}. For isotropic systems $\Delta \hat{\Upsilon}_n=0$ for even orders, thus in figure \ref{fig:IsoMSP34} we show the non-trivial MSP, i.e. the third order. The simulated results presented in this section have been obtained using the methodology presented in this paper for random systems. These structures have been generated using the BRF method with exponential, Gaussian and slightly damped sinc kernels. The simulation parameters used are as follows: the material permittivity $\epsilon_1=10$ and the void phase permittivity $\epsilon_2=1$, we performed an ensemble average of 100 realizations, and the ratio between the decay length of the kernel and the unit cell is 1/30.  The sinc kernel was slightly damped with a 10 times slower Gaussian decay to ensure sufficient decay while retaining short to medium range oscillation.

First, we consider how the third order MSP of isotropic kernels is affected by fill factor ($\phi=0.1\rightarrow0.9$), as seen in figure \ref{fig:IsoMSP34}.  As expected our results show that the structure at 50 $\%$ fill-factor has $<\hat{\Upsilon}_3>=1/2$. It is interesting to compare the third order of each kernel with the fill-factor (green solid line), since the third order at low and high fill-factors is associated with the shape of nearly-isolated inclusions. We can see the third order is closest to first order for the sinc kernel, while for the Gaussian and exponential kernels the orders diverge. The superior size homogeneity of the sinc structure seen in figure \ref{fig:IsoMSP34} leads us to conjecture that the third order parameter is correlated with size homogeneity in these structures.

Next we analyze the anisotropy as represented by the different in each even order MSP tensor (see equation \ref{eq:DeltUps}). In figure \ref{fig:AniMSP34} we show second and fourth order MSP for different kernel structures with a fixed fill-factor of 50 $\%$, and varying the stretch ratio from $e_y/e_x=1$ (isotropic) to $e_y/e_x=0.2$. In the left panel of figure \ref{fig:AniMSP34}, we can see that second order MSP of the three different structures converges to the depolarization of an ellipse with the same anisotropy ratio (note that without damping the convergence of the sinc is slower). This result agrees with the explanation given above.  The right panel in figure \ref{fig:AniMSP34} shows $\Delta \hat{\Upsilon}_4$. All of the kernels have a small positive difference of the fourth order parameter: inspection shows that this will act to reduce the anisotropy of the permittivity in the context of a negative difference of the second order parameter.  We can see that sinc kernel structures offer somewhat greater values of $\Delta \hat{\Upsilon}_4$ than Gaussian or exponential kernels, but this difference should only have a minor effect unless the permittivity contrast is large.

\subsection{Macroscopic Permittivity}
We also examined the difference between the diagonal elements of the effective permittivity of each system (exponential, Gaussian and damped sinc kernel structures), for fill factor 1/2, and inclusion permittivity 10 (this is reasonably representative for high index dielectrics). We found that the main effect of the anisotropy of the permittivity is due to the anisotropy ratio of the unit cell, and that the choice of generating kernel has a small effect on the total anisotropy for the permittivity chosen here.  At aspect ratio 0.2 we found a difference of about +/-0.03 between kernels compared to an average of about 2.3.  Careful inspection shows that the exponential results have the highest anisotropy consistent with the weakest opposing fourth order parameter.  At higher or lower fill factors, the absolute effect of high orders will be greatly reduced, so we expect that the effective anisotropy is dominated by second order for all fill factors at low contrast.  At higher permittivity contrast, the higher order parameters will become more important.

\section{Conclusion}
Building on previous work \cite{Brown1955,Milton1981,Milton2002,Torquato1985a,Torquato2002}, we have developed a generalized framework for predicting the effective electrical parameters of composites, that does not require the explicit evaluation of correlations or multidimensional integrals. This facilitates the extraction of microstructural parameters and a series of tight bounds on the effective anisotropic parameters. This framework can be applied to high orders with the caveat that numerical sampling requirements become more stringent.  We demonstrated the utility of this framework in predicting the properties of two-dimensional level-cut random fields. We found that the depolarization of elliptical kernels is the same as equivalent ellipses, and this result should extend to ellipsoids (consistent with a similar conclusion \citet{Hori1973,Hori1973a} for the less correlated symmetric-cell structures). Third order results demonstrate the expected gradual increase as they pass through 0.5 at percolation (50 $\%$ fill factor), with the sinc kernel closest to linearity, while the exponential and Gaussian kernels at fill-factors appear to have some residual third order at the extremes. We have found that the choice of kernel has only a minor effect on fourth order, with a minor opposing effect which was largest for the sinc kernel. Finally, the anisotropy in the macroscopic permittivity is primarily controlled by the stretch factor of the unit cell, and the detail of the kernel distributions chosen here has negligible effect at contrasts applicable to typical optical dielectrics.

% Bibliography
%\section*{References}
% Bibliography: source file
\bibliography{ColumnsPaperF.bib}

\begin{thebibliography}{38}%
\makeatletter
\providecommand \@ifxundefined [1]{%
 \@ifx{#1\undefined}
}%
\providecommand \@ifnum [1]{%
 \ifnum #1\expandafter \@firstoftwo
 \else \expandafter \@secondoftwo
 \fi
}%
\providecommand \@ifx [1]{%
 \ifx #1\expandafter \@firstoftwo
 \else \expandafter \@secondoftwo
 \fi
}%
\providecommand \natexlab [1]{#1}%
\providecommand \enquote  [1]{``#1''}%
\providecommand \bibnamefont  [1]{#1}%
\providecommand \bibfnamefont [1]{#1}%
\providecommand \citenamefont [1]{#1}%
\providecommand \href@noop [0]{\@secondoftwo}%
\providecommand \href [0]{\begingroup \@sanitize@url \@href}%
\providecommand \@href[1]{\@@startlink{#1}\@@href}%
\providecommand \@@href[1]{\endgroup#1\@@endlink}%
\providecommand \@sanitize@url [0]{\catcode `\\12\catcode `\$12\catcode
  `\&12\catcode `\#12\catcode `\^12\catcode `\_12\catcode `\%12\relax}%
\providecommand \@@startlink[1]{}%
\providecommand \@@endlink[0]{}%
\providecommand \url  [0]{\begingroup\@sanitize@url \@url }%
\providecommand \@url [1]{\endgroup\@href {#1}{\urlprefix }}%
\providecommand \urlprefix  [0]{URL }%
\providecommand \Eprint [0]{\href }%
\providecommand \doibase [0]{http://dx.doi.org/}%
\providecommand \selectlanguage [0]{\@gobble}%
\providecommand \bibinfo  [0]{\@secondoftwo}%
\providecommand \bibfield  [0]{\@secondoftwo}%
\providecommand \translation [1]{[#1]}%
\providecommand \BibitemOpen [0]{}%
\providecommand \bibitemStop [0]{}%
\providecommand \bibitemNoStop [0]{.\EOS\space}%
\providecommand \EOS [0]{\spacefactor3000\relax}%
\providecommand \BibitemShut  [1]{\csname bibitem#1\endcsname}%
\let\auto@bib@innerbib\@empty
%</preamble>
\bibitem [{\citenamefont {Brown}(1955)}]{Brown1955}%
  \BibitemOpen
  \bibfield  {author} {\bibinfo {author} {\bibfnamefont {W.~F.}\ \bibnamefont
  {Brown}},\ }\href {\doibase 10.1063/1.1742339} {\bibfield  {journal}
  {\bibinfo  {journal} {The Journal of Chemical Physics}\ }\textbf {\bibinfo
  {volume} {23}},\ \bibinfo {pages} {1514} (\bibinfo {year}
  {1955})}\BibitemShut {NoStop}%
\bibitem [{\citenamefont {Milton}(1981)}]{Milton1981}%
  \BibitemOpen
  \bibfield  {author} {\bibinfo {author} {\bibfnamefont {G.~W.}\ \bibnamefont
  {Milton}},\ }\href {\doibase 10.1063/1.329385} {\bibfield  {journal}
  {\bibinfo  {journal} {Journal of Applied Physics}\ }\textbf {\bibinfo
  {volume} {52}},\ \bibinfo {pages} {5286} (\bibinfo {year}
  {1981})}\BibitemShut {NoStop}%
\bibitem [{\citenamefont {Milton}(2002)}]{Milton2002}%
  \BibitemOpen
  \bibfield  {author} {\bibinfo {author} {\bibfnamefont {G.~W.}\ \bibnamefont
  {Milton}},\ }\href@noop {} {\emph {\bibinfo {title} {{The Theory of
  Composites}}}}\ (\bibinfo  {publisher} {Cambridge University Press},\
  \bibinfo {year} {2002})\BibitemShut {NoStop}%
\bibitem [{\citenamefont {Torquato}(1985)}]{Torquato1985a}%
  \BibitemOpen
  \bibfield  {author} {\bibinfo {author} {\bibfnamefont {S.}~\bibnamefont
  {Torquato}},\ }\href {\doibase 10.1063/1.335593} {\bibfield  {journal}
  {\bibinfo  {journal} {Journal of Applied Physics}\ }\textbf {\bibinfo
  {volume} {58}},\ \bibinfo {pages} {3790} (\bibinfo {year}
  {1985})}\BibitemShut {NoStop}%
\bibitem [{\citenamefont {Torquato}(2002)}]{Torquato2002}%
  \BibitemOpen
  \bibfield  {author} {\bibinfo {author} {\bibfnamefont {S.}~\bibnamefont
  {Torquato}},\ }\href@noop {} {\emph {\bibinfo {title} {{Random Heterogeneous
  Materials: Microstructure and Macroscopic Properties}}}}\ (\bibinfo
  {publisher} {Springer-Verlag},\ \bibinfo {address} {New York},\ \bibinfo
  {year} {2002})\BibitemShut {NoStop}%
\bibitem [{\citenamefont {Walsby}\ \emph {et~al.}(2005)\citenamefont {Walsby},
  \citenamefont {Arnold}, \citenamefont {Wu}, \citenamefont {Hodgkinson},\ and\
  \citenamefont {Blaikie}}]{Walsby2005}%
  \BibitemOpen
  \bibfield  {author} {\bibinfo {author} {\bibfnamefont {E.~D.}\ \bibnamefont
  {Walsby}}, \bibinfo {author} {\bibfnamefont {M.}~\bibnamefont {Arnold}},
  \bibinfo {author} {\bibfnamefont {Q.~H.}\ \bibnamefont {Wu}}, \bibinfo
  {author} {\bibfnamefont {I.~J.}\ \bibnamefont {Hodgkinson}}, \ and\ \bibinfo
  {author} {\bibfnamefont {R.~J.}\ \bibnamefont {Blaikie}},\ }\href {\doibase
  10.1016/j.mee.2004.12.055} {\bibfield  {journal} {\bibinfo  {journal}
  {Microelectronic Engineering}\ }\textbf {\bibinfo {volume} {78-79}},\
  \bibinfo {pages} {436} (\bibinfo {year} {2005})}\BibitemShut {NoStop}%
\bibitem [{\citenamefont {Tai}\ \emph {et~al.}(2017)\citenamefont {Tai},
  \citenamefont {Arnold}, \citenamefont {Gentle},\ and\ \citenamefont
  {Smith}}]{Tai2017}%
  \BibitemOpen
  \bibfield  {author} {\bibinfo {author} {\bibfnamefont {M.~C.}\ \bibnamefont
  {Tai}}, \bibinfo {author} {\bibfnamefont {M.~D.}\ \bibnamefont {Arnold}},
  \bibinfo {author} {\bibfnamefont {A.~R.}\ \bibnamefont {Gentle}}, \ and\
  \bibinfo {author} {\bibfnamefont {G.~B.}\ \bibnamefont {Smith}},\ }in\ \href
  {\doibase 10.1117/12.2273528} {\emph {\bibinfo {booktitle} {Nanostructured
  Thin Films X}}},\ \bibinfo {editor} {edited by\ \bibinfo {editor}
  {\bibfnamefont {T.~G.}\ \bibnamefont {Mackay}}, \bibinfo {editor}
  {\bibfnamefont {A.}~\bibnamefont {Lakhtakia}}, \ and\ \bibinfo {editor}
  {\bibfnamefont {Y.-J.}\ \bibnamefont {Jen}}}\ (\bibinfo  {publisher} {SPIE},\
  \bibinfo {year} {2017})\ p.~\bibinfo {pages} {20}\BibitemShut {NoStop}%
\bibitem [{\citenamefont {Teubner}(1991)}]{Teubner1991}%
  \BibitemOpen
  \bibfield  {author} {\bibinfo {author} {\bibfnamefont {M.}~\bibnamefont
  {Teubner}},\ }\href@noop {} {\bibfield  {journal} {\bibinfo  {journal}
  {Europhysics Letters (EPL)}\ }\textbf {\bibinfo {volume} {14}},\ \bibinfo
  {pages} {403} (\bibinfo {year} {1991})}\BibitemShut {NoStop}%
\bibitem [{\citenamefont {Sen}\ and\ \citenamefont {Torquato}(1989)}]{Sen1989}%
  \BibitemOpen
  \bibfield  {author} {\bibinfo {author} {\bibfnamefont {A.~K.}\ \bibnamefont
  {Sen}}\ and\ \bibinfo {author} {\bibfnamefont {S.}~\bibnamefont {Torquato}},\
  }\href@noop {} {\bibfield  {journal} {\bibinfo  {journal} {Physical Review
  B}\ }\textbf {\bibinfo {volume} {39}},\ \bibinfo {pages} {4504} (\bibinfo
  {year} {1989})}\BibitemShut {NoStop}%
\bibitem [{\citenamefont {Miller}(1969{\natexlab{a}})}]{Miller1969}%
  \BibitemOpen
  \bibfield  {author} {\bibinfo {author} {\bibfnamefont {M.~N.}\ \bibnamefont
  {Miller}},\ }\href {\doibase 10.1063/1.1664794} {\bibfield  {journal}
  {\bibinfo  {journal} {Journal of Mathematical Physics}\ }\textbf {\bibinfo
  {volume} {10}},\ \bibinfo {pages} {1988} (\bibinfo {year}
  {1969}{\natexlab{a}})}\BibitemShut {NoStop}%
\bibitem [{\citenamefont {Bergman}(1981)}]{Bergman1981}%
  \BibitemOpen
  \bibfield  {author} {\bibinfo {author} {\bibfnamefont {D.~J.}\ \bibnamefont
  {Bergman}},\ }\href {\doibase 10.1063/1.91895} {\bibfield  {journal}
  {\bibinfo  {journal} {Phys. Rev. B}\ }\textbf {\bibinfo {volume} {23}},\
  \bibinfo {pages} {3058} (\bibinfo {year} {1981})}\BibitemShut {NoStop}%
\bibitem [{\citenamefont {Milton}(1982)}]{Milton1982}%
  \BibitemOpen
  \bibfield  {author} {\bibinfo {author} {\bibfnamefont {G.~W.}\ \bibnamefont
  {Milton}},\ }\href@noop {} {\bibfield  {journal} {\bibinfo  {journal}
  {Journal of the Mechanics and Physics of Solids}\ }\textbf {\bibinfo {volume}
  {30}},\ \bibinfo {pages} {177} (\bibinfo {year} {1982})}\BibitemShut
  {NoStop}%
\bibitem [{\citenamefont {Torquato}\ and\ \citenamefont
  {Lado}(1991)}]{Torquato1991}%
  \BibitemOpen
  \bibfield  {author} {\bibinfo {author} {\bibfnamefont {S.}~\bibnamefont
  {Torquato}}\ and\ \bibinfo {author} {\bibfnamefont {F.}~\bibnamefont
  {Lado}},\ }\href {\doibase 10.1063/1.460635} {\bibfield  {journal} {\bibinfo
  {journal} {The Journal of Chemical Physics}\ }\textbf {\bibinfo {volume}
  {94}},\ \bibinfo {pages} {4453} (\bibinfo {year} {1991})}\BibitemShut
  {NoStop}%
\bibitem [{\citenamefont {Torquato}\ and\ \citenamefont
  {Stell}(1983)}]{Torquato1983}%
  \BibitemOpen
  \bibfield  {author} {\bibinfo {author} {\bibfnamefont {S.}~\bibnamefont
  {Torquato}}\ and\ \bibinfo {author} {\bibfnamefont {G.}~\bibnamefont
  {Stell}},\ }\href {\doibase 10.1063/1.445941} {\bibfield  {journal} {\bibinfo
   {journal} {The Journal of Chemical Physics}\ }\textbf {\bibinfo {volume}
  {79}},\ \bibinfo {pages} {1505} (\bibinfo {year} {1983})}\BibitemShut
  {NoStop}%
\bibitem [{\citenamefont {Torquato}\ and\ \citenamefont
  {Sen}(1990)}]{Torquato1990}%
  \BibitemOpen
  \bibfield  {author} {\bibinfo {author} {\bibfnamefont {S.}~\bibnamefont
  {Torquato}}\ and\ \bibinfo {author} {\bibfnamefont {A.~K.}\ \bibnamefont
  {Sen}},\ }\href {\doibase 10.1063/1.345711} {\bibfield  {journal} {\bibinfo
  {journal} {Journal of Applied Physics}\ }\textbf {\bibinfo {volume} {67}},\
  \bibinfo {pages} {1145} (\bibinfo {year} {1990})}\BibitemShut {NoStop}%
\bibitem [{\citenamefont {Cherkaev}\ and\ \citenamefont
  {Gibiansky}(1996)}]{Cherkaev1996}%
  \BibitemOpen
  \bibfield  {author} {\bibinfo {author} {\bibfnamefont {A.~V.}\ \bibnamefont
  {Cherkaev}}\ and\ \bibinfo {author} {\bibfnamefont {L.~V.}\ \bibnamefont
  {Gibiansky}},\ }\href {\doibase 10.1016/0020-7683(95)00176-X} {\bibfield
  {journal} {\bibinfo  {journal} {International Journal of Solids and
  Structures}\ }\textbf {\bibinfo {volume} {33}},\ \bibinfo {pages} {2609}
  (\bibinfo {year} {1996})}\BibitemShut {NoStop}%
\bibitem [{\citenamefont {Wiener}(1912)}]{Wiener1912}%
  \BibitemOpen
  \bibfield  {author} {\bibinfo {author} {\bibfnamefont {O.}~\bibnamefont
  {Wiener}},\ }\href@noop {} {\bibfield  {journal} {\bibinfo  {journal} {Abh.
  Math. Phys. Kl. Kgl. Sachs. Ges.}\ }\textbf {\bibinfo {volume} {32}},\
  \bibinfo {pages} {509} (\bibinfo {year} {1912})}\BibitemShut {NoStop}%
\bibitem [{\citenamefont {Murat}\ and\ \citenamefont
  {Tartar}(1994)}]{Murat1994}%
  \BibitemOpen
  \bibfield  {author} {\bibinfo {author} {\bibfnamefont {F.}~\bibnamefont
  {Murat}}\ and\ \bibinfo {author} {\bibfnamefont {L.}~\bibnamefont {Tartar}},\
  }\href@noop {} {\emph {\bibinfo {title} {{On the Control of Coefficients in
  Partial Differential Equations}}}}\ (\bibinfo  {publisher}
  {Birkh{\"{a}}user},\ \bibinfo {address} {Boston, MA},\ \bibinfo {year}
  {1994})\BibitemShut {NoStop}%
\bibitem [{\citenamefont {Engstr{\"{o}}m}(2005)}]{Engstrom2005}%
  \BibitemOpen
  \bibfield  {author} {\bibinfo {author} {\bibfnamefont {C.}~\bibnamefont
  {Engstr{\"{o}}m}},\ }\href {\doibase 10.1088/0022-3727/38/19/019} {\bibfield
  {journal} {\bibinfo  {journal} {Journal of Physics D: Applied Physics}\
  }\textbf {\bibinfo {volume} {38}},\ \bibinfo {pages} {3695} (\bibinfo {year}
  {2005})}\BibitemShut {NoStop}%
\bibitem [{\citenamefont {Arnold}(2017)}]{Arnold2017}%
  \BibitemOpen
  \bibfield  {author} {\bibinfo {author} {\bibfnamefont {M.~D.}\ \bibnamefont
  {Arnold}},\ }\href {\doibase 10.1088/1361-648X/aa57c8} {\bibfield  {journal}
  {\bibinfo  {journal} {Journal of Physics Condensed Matter}\ }\textbf
  {\bibinfo {volume} {29}} (\bibinfo {year} {2017}),\
  10.1088/1361-648X/aa57c8}\BibitemShut {NoStop}%
\bibitem [{\citenamefont {Al{\`{u}}}(2011)}]{Alu2011}%
  \BibitemOpen
  \bibfield  {author} {\bibinfo {author} {\bibfnamefont {A.}~\bibnamefont
  {Al{\`{u}}}},\ }\href {\doibase 10.1103/PhysRevB.84.075153} {\bibfield
  {journal} {\bibinfo  {journal} {Physical Review B - Condensed Matter and
  Materials Physics}\ }\textbf {\bibinfo {volume} {84}},\ \bibinfo {pages} {1}
  (\bibinfo {year} {2011})},\ \Eprint {http://arxiv.org/abs/1012.1351}
  {arXiv:1012.1351} \BibitemShut {NoStop}%
\bibitem [{\citenamefont {Simovski}\ \emph {et~al.}(2012)\citenamefont
  {Simovski}, \citenamefont {Belov}, \citenamefont {Atrashchenko},\ and\
  \citenamefont {Kivshar}}]{Simovski2012}%
  \BibitemOpen
  \bibfield  {author} {\bibinfo {author} {\bibfnamefont {C.~R.}\ \bibnamefont
  {Simovski}}, \bibinfo {author} {\bibfnamefont {P.~A.}\ \bibnamefont {Belov}},
  \bibinfo {author} {\bibfnamefont {A.~V.}\ \bibnamefont {Atrashchenko}}, \
  and\ \bibinfo {author} {\bibfnamefont {Y.~S.}\ \bibnamefont {Kivshar}},\
  }\href {\doibase 10.1002/adma.201200931} {\bibfield  {journal} {\bibinfo
  {journal} {Advanced Materials}\ }\textbf {\bibinfo {volume} {24}},\ \bibinfo
  {pages} {4229} (\bibinfo {year} {2012})}\BibitemShut {NoStop}%
\bibitem [{\citenamefont {Draine}\ and\ \citenamefont
  {Flatau}(1994)}]{Draine1994}%
  \BibitemOpen
  \bibfield  {author} {\bibinfo {author} {\bibfnamefont {B.~T.}\ \bibnamefont
  {Draine}}\ and\ \bibinfo {author} {\bibfnamefont {P.~J.}\ \bibnamefont
  {Flatau}},\ }\href {\doibase 10.1364/JOSAA.11.001491} {\bibfield  {journal}
  {\bibinfo  {journal} {Journal of the Optical Society of America A}\ }\textbf
  {\bibinfo {volume} {11}},\ \bibinfo {pages} {1491} (\bibinfo {year}
  {1994})}\BibitemShut {NoStop}%
\bibitem [{\citenamefont {Draine}\ and\ \citenamefont
  {Flatau}(2008)}]{Draine2008}%
  \BibitemOpen
  \bibfield  {author} {\bibinfo {author} {\bibfnamefont {B.~T.}\ \bibnamefont
  {Draine}}\ and\ \bibinfo {author} {\bibfnamefont {P.~J.}\ \bibnamefont
  {Flatau}},\ }\href {\doibase 10.1364/JOSAA.25.002693} {\bibfield  {journal}
  {\bibinfo  {journal} {Journal of the Optical Society of America. A, Optics,
  image science, and vision}\ }\textbf {\bibinfo {volume} {25}},\ \bibinfo
  {pages} {2693} (\bibinfo {year} {2008})},\ \Eprint
  {http://arxiv.org/abs/0809.0338} {arXiv:0809.0338} \BibitemShut {NoStop}%
\bibitem [{\citenamefont {O'Donnell}(2001)}]{ODonnell2001}%
  \BibitemOpen
  \bibfield  {author} {\bibinfo {author} {\bibfnamefont {K.~A.}\ \bibnamefont
  {O'Donnell}},\ }\href {\doibase 10.1364/JOSAA.18.001507} {\bibfield
  {journal} {\bibinfo  {journal} {Journal of the Optical Society of America A}\
  }\textbf {\bibinfo {volume} {18}},\ \bibinfo {pages} {1507} (\bibinfo {year}
  {2001})}\BibitemShut {NoStop}%
\bibitem [{\citenamefont {Hashin}\ and\ \citenamefont
  {Shtrikman}(1962)}]{Hashin1962}%
  \BibitemOpen
  \bibfield  {author} {\bibinfo {author} {\bibfnamefont {Z.}~\bibnamefont
  {Hashin}}\ and\ \bibinfo {author} {\bibfnamefont {S.}~\bibnamefont
  {Shtrikman}},\ }\href {\doibase 10.1063/1.1728579} {\bibfield  {journal}
  {\bibinfo  {journal} {Journal of Applied Physics}\ }\textbf {\bibinfo
  {volume} {33}},\ \bibinfo {pages} {3125} (\bibinfo {year}
  {1962})}\BibitemShut {NoStop}%
\bibitem [{\citenamefont {Keller}(1964)}]{Keller1964}%
  \BibitemOpen
  \bibfield  {author} {\bibinfo {author} {\bibfnamefont {J.~B.}\ \bibnamefont
  {Keller}},\ }\href {\doibase 10.1063/1.1704146} {\bibfield  {journal}
  {\bibinfo  {journal} {Journal of Mathematical Physics}\ }\textbf {\bibinfo
  {volume} {5}},\ \bibinfo {pages} {548} (\bibinfo {year} {1964})}\BibitemShut
  {NoStop}%
\bibitem [{\citenamefont {Hetherington}\ and\ \citenamefont
  {Thorpe}(1992)}]{Hetherington1992}%
  \BibitemOpen
  \bibfield  {author} {\bibinfo {author} {\bibfnamefont {J.~H.}\ \bibnamefont
  {Hetherington}}\ and\ \bibinfo {author} {\bibfnamefont {M.~F.}\ \bibnamefont
  {Thorpe}},\ }\href@noop {} {\bibfield  {journal} {\bibinfo  {journal}
  {Proceedings: Mathematical and Physical Scences}\ }\textbf {\bibinfo {volume}
  {438}},\ \bibinfo {pages} {591} (\bibinfo {year} {1992})}\BibitemShut
  {NoStop}%
\bibitem [{\citenamefont {McPhedran}\ and\ \citenamefont
  {Milton}(1981)}]{McPhedran1981}%
  \BibitemOpen
  \bibfield  {author} {\bibinfo {author} {\bibfnamefont {R.~C.}\ \bibnamefont
  {McPhedran}}\ and\ \bibinfo {author} {\bibfnamefont {G.~W.}\ \bibnamefont
  {Milton}},\ }\href {\doibase 10.1007/BF00617840} {\bibfield  {journal}
  {\bibinfo  {journal} {Applied Physics A Solids and Surfaces}\ }\textbf
  {\bibinfo {volume} {26}},\ \bibinfo {pages} {207} (\bibinfo {year}
  {1981})}\BibitemShut {NoStop}%
\bibitem [{\citenamefont {Yardley}\ \emph {et~al.}(1999)\citenamefont
  {Yardley}, \citenamefont {McPhedran}, \citenamefont {Nicorovici},\ and\
  \citenamefont {Botten}}]{Yardley1999}%
  \BibitemOpen
  \bibfield  {author} {\bibinfo {author} {\bibfnamefont {J.~G.}\ \bibnamefont
  {Yardley}}, \bibinfo {author} {\bibfnamefont {R.~C.}\ \bibnamefont
  {McPhedran}}, \bibinfo {author} {\bibfnamefont {N.~A.}\ \bibnamefont
  {Nicorovici}}, \ and\ \bibinfo {author} {\bibfnamefont {L.~C.}\ \bibnamefont
  {Botten}},\ }\href {\doibase 10.1103/PhysRevE.60.6068} {\bibfield  {journal}
  {\bibinfo  {journal} {Physical Review E - Statistical Physics, Plasmas,
  Fluids, and Related Interdisciplinary Topics}\ }\textbf {\bibinfo {volume}
  {60}},\ \bibinfo {pages} {6068} (\bibinfo {year} {1999})}\BibitemShut
  {NoStop}%
\bibitem [{\citenamefont {Lu}(1999)}]{Lu1999}%
  \BibitemOpen
  \bibfield  {author} {\bibinfo {author} {\bibfnamefont {S.~Y.}\ \bibnamefont
  {Lu}},\ }\href {\doibase 10.1063/1.369439} {\bibfield  {journal} {\bibinfo
  {journal} {Journal of Applied Physics}\ }\textbf {\bibinfo {volume} {85}},\
  \bibinfo {pages} {264} (\bibinfo {year} {1999})}\BibitemShut {NoStop}%
\bibitem [{\citenamefont {McPhedran}\ and\ \citenamefont
  {McKenzie}(1980)}]{McPhedran1980}%
  \BibitemOpen
  \bibfield  {author} {\bibinfo {author} {\bibfnamefont {R.~C.}\ \bibnamefont
  {McPhedran}}\ and\ \bibinfo {author} {\bibfnamefont {D.~R.}\ \bibnamefont
  {McKenzie}},\ }\href {\doibase 10.1093/rpd/ncx264} {\bibfield  {journal}
  {\bibinfo  {journal} {Applied Physics}\ }\textbf {\bibinfo {volume} {23}},\
  \bibinfo {pages} {223} (\bibinfo {year} {1980})}\BibitemShut {NoStop}%
\bibitem [{\citenamefont {Nicorovici}\ and\ \citenamefont
  {McPhedran}(1996)}]{Nicorovici1996}%
  \BibitemOpen
  \bibfield  {author} {\bibinfo {author} {\bibfnamefont {N.~A.}\ \bibnamefont
  {Nicorovici}}\ and\ \bibinfo {author} {\bibfnamefont {R.~C.}\ \bibnamefont
  {McPhedran}},\ }\href {http://www.ncbi.nlm.nih.gov/pubmed/9965278} {\bibfield
   {journal} {\bibinfo  {journal} {Physical review. E, Statistical physics,
  plasmas, fluids, and related interdisciplinary topics}\ }\textbf {\bibinfo
  {volume} {54}},\ \bibinfo {pages} {1945} (\bibinfo {year}
  {1996})}\BibitemShut {NoStop}%
\bibitem [{\citenamefont {Roberts}\ and\ \citenamefont
  {Teubner}(1995)}]{Roberts1995}%
  \BibitemOpen
  \bibfield  {author} {\bibinfo {author} {\bibfnamefont {A.~P.}\ \bibnamefont
  {Roberts}}\ and\ \bibinfo {author} {\bibfnamefont {M.}~\bibnamefont
  {Teubner}},\ }\href@noop {} {\bibfield  {journal} {\bibinfo  {journal}
  {Physical Review E}\ }\textbf {\bibinfo {volume} {51}},\ \bibinfo {pages}
  {4141} (\bibinfo {year} {1995})}\BibitemShut {NoStop}%
\bibitem [{\citenamefont {Roberts}\ and\ \citenamefont
  {Torquato}(1999)}]{Roberts1999}%
  \BibitemOpen
  \bibfield  {author} {\bibinfo {author} {\bibfnamefont {A.~P.}\ \bibnamefont
  {Roberts}}\ and\ \bibinfo {author} {\bibfnamefont {S.}~\bibnamefont
  {Torquato}},\ }\href {\doibase 10.1103/PhysRevE.59.4953} {\bibfield
  {journal} {\bibinfo  {journal} {Physical Review E - Statistical Physics,
  Plasmas, Fluids, and Related Interdisciplinary Topics}\ }\textbf {\bibinfo
  {volume} {59}},\ \bibinfo {pages} {4953} (\bibinfo {year} {1999})},\ \Eprint
  {http://arxiv.org/abs/9901315} {arXiv:9901315 [cond-mat]} \BibitemShut
  {NoStop}%
\bibitem [{\citenamefont {Miller}(1969{\natexlab{b}})}]{Miller1969b}%
  \BibitemOpen
  \bibfield  {author} {\bibinfo {author} {\bibfnamefont {M.~N.}\ \bibnamefont
  {Miller}},\ }\href {\doibase 10.1063/1.1664795} {\bibfield  {journal}
  {\bibinfo  {journal} {Journal of Mathematical Physics}\ }\textbf {\bibinfo
  {volume} {10}},\ \bibinfo {pages} {2005} (\bibinfo {year}
  {1969}{\natexlab{b}})}\BibitemShut {NoStop}%
\bibitem [{\citenamefont {Hori}(1973{\natexlab{a}})}]{Hori1973}%
  \BibitemOpen
  \bibfield  {author} {\bibinfo {author} {\bibfnamefont {M.}~\bibnamefont
  {Hori}},\ }\href {\doibase 10.1063/1.1666275} {\bibfield  {journal} {\bibinfo
   {journal} {Journal of Mathematical Physics}\ }\textbf {\bibinfo {volume}
  {14}},\ \bibinfo {pages} {1942} (\bibinfo {year}
  {1973}{\natexlab{a}})}\BibitemShut {NoStop}%
\bibitem [{\citenamefont {Hori}(1973{\natexlab{b}})}]{Hori1973a}%
  \BibitemOpen
  \bibfield  {author} {\bibinfo {author} {\bibfnamefont {M.}~\bibnamefont
  {Hori}},\ }\href {\doibase 10.1063/1.1666347} {\bibfield  {journal} {\bibinfo
   {journal} {Journal of Mathematical Physics}\ }\textbf {\bibinfo {volume}
  {14}},\ \bibinfo {pages} {514} (\bibinfo {year}
  {1973}{\natexlab{b}})}\BibitemShut {NoStop}%
\end{thebibliography}%

\end{document}